\documentclass[10pt]{iopart}
\usepackage{iopams}
\usepackage{epsfig}
\usepackage{amssymb}
\usepackage{mathrsfs}

\begin{document}

\title[Saitoh, Koshiba \& Mortensen:
Nonlinear photonic crystal fibres ...]{Nonlinear photonic crystal
fibres: pushing the zero-dispersion toward the visible }

\author{Kunimasa Saitoh and Masanori Koshiba}

\address{Hokkaido University, North 14 West 9, Kita-ku, Sapporo, 060-0814, Japan}

\author{Niels Asger Mortensen\footnote[3]{Corresponding author: nam@mic.dtu.dk}}

\address{MIC -- Department of Micro and Nanotechnology, NanoDTU,
Technical University of Denmark, Bld. 345 east, DK-2800 Kongens Lyngby, Denmark}

\begin{abstract}
The strong waveguide dispersion in photonic crystal fibres
provides unique opportunities for nonlinear optics with a
zero-dispersion wavelength $\lambda_0$ far below the limit of
$\sim 1.3\,{\rm \mu m}$ set by the material dispersion of silica.
By tuning the air-hole diameter $d$, the pitch $\Lambda$, and the
number of rings of air holes $N$, the strong waveguide dispersion
can in principle be used to extend $\lambda_0$ well into the
visible, albeit to some extend at the cost of multimode operation.
We study in detail the interplay of the zero-dispersion
wavelength, the cut-off wavelength $\lambda_c$, and the leakage
loss in the parameter space spanned by $d$, $\Lambda$, and $N$. As
a particular result we identify values of $d$ ($\sim 500$~nm) and
$\Lambda$ ($\sim 700$~nm) which facilitate the shortest possible
zero-dispersion wavelength ($\sim 700$~nm) while the fibre is
still single-mode for longer wavelengths.
\end{abstract}


\submitto{\NJP (Focus on Nanophotonics)}

 \maketitle

\section{Introduction}

Photonic crystal fibres (PCFs)~\cite{Russell:2003,Knight:2003}
have led to an enormous renewed interest in nonlinear fibre
optics~\cite{Mollenauer:2003,Hansen:2005,Zheltikov:2006,Zheltikov:2006b}.
In particular, the strong anomalous group-velocity dispersion has
facilitated visible super-continuum generation~\cite{Ranka:2000},
which apart from being a fascinating and rich non-linear
phenomenon~\cite{Herrmann:2002} also has promising applications
including frequency-comb metrology~\cite{Jones:2000,Udem:2002} and
optical coherence tomography~\cite{Hartl:2001}.

\begin{figure}[b]
\begin{center}
\epsfig{file=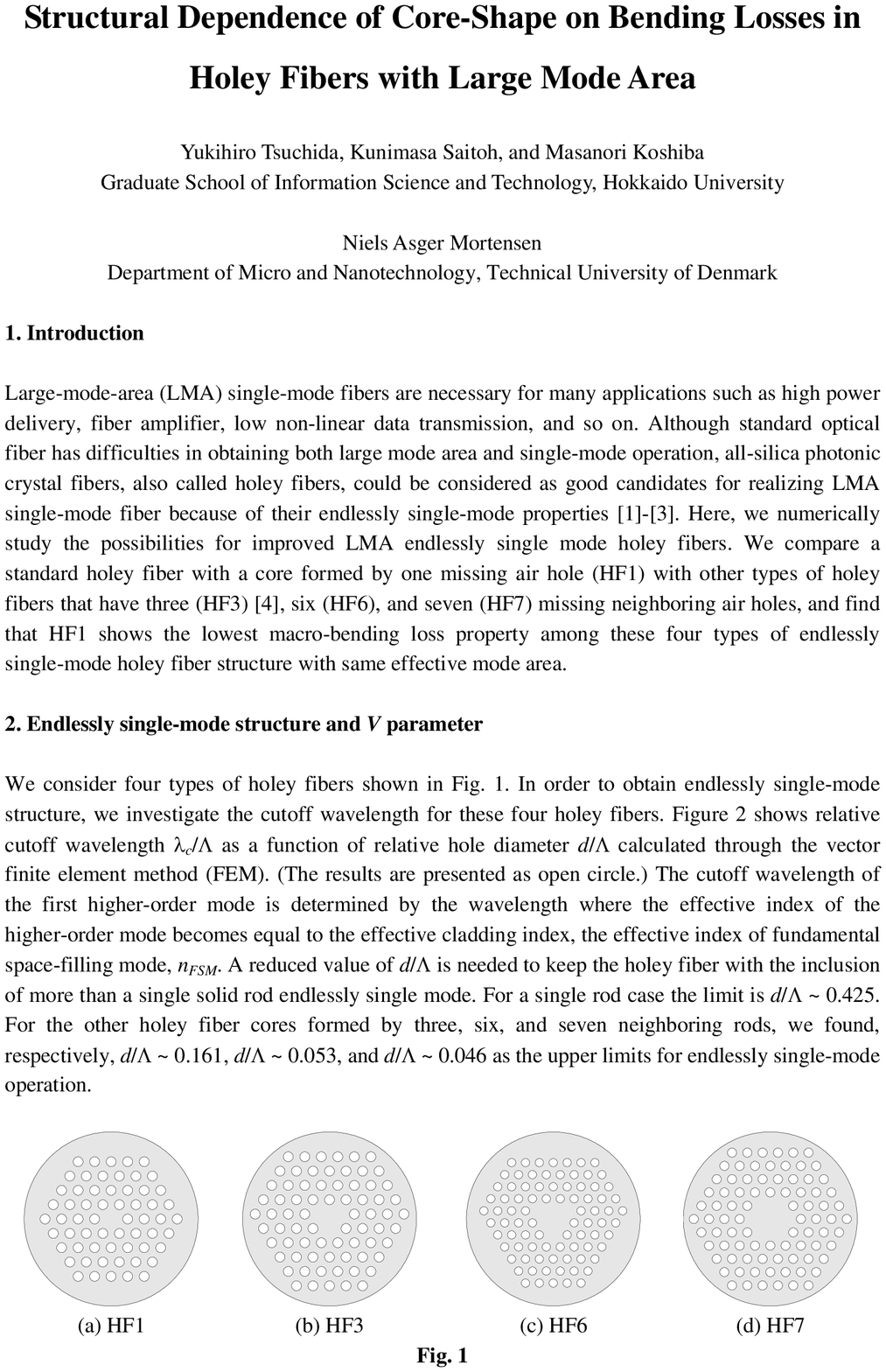, width=0.45\textwidth,clip}
\end{center}
\caption{Cross section of a photonic crystal fibre with $N=4$
rings of air holes in the photonic crystal cladding surrounding
the core defect which is formed by the omission of a single air
hole in the otherwise periodic structure. The photonic crystal
cladding comprises air holes of diameter $d$ arranged in a
triangular array of pitch $\Lambda$. } \label{fig1}
\end{figure}

The regular triangular arrangement of sub-micron air-holes running
along the full length of the fibre \cite{Knight:1996,Birks:1997},
see Fig.~\ref{fig1}, is a key concept for the realization of
strong anomalous chromatic dispersion which arises in a
competition between material dispersion and wave-guide dispersion
originating from the strong transverse confinement of light. At
the same time the strong transverse confinement of light also
serves to dramatically decrease the effective area and increase
the non-linear coefficient~\cite{Mortensen:2002} so that very high
optical intensities may be achieved relative to the input power.

The zero-dispersion wavelength may be tuned down to the visible
\cite{Knight:2000,Skryabin:2003} by carefully increasing the
normalized air-hole diameter $d/\Lambda$ while at the same time
decreasing the pitch $\Lambda$. However, in many cases the very
short zero-dispersion wavelength is achieved at the cost of
multi-mode operation. Furthermore, the reduction of the pitch may
require a large number $N$ of rings of air holes in order to
circumvent leakage loss.

In this paper we in detail study the complicated interplay of the
zero-dispersion wavelength $\lambda_0$, the cut-off wavelength
$\lambda_c$, and the leakage loss in the parameter space spanned
by $d$, $\Lambda$, and $N$. As a particular result we identify
values of $d$ and $\Lambda$ which facilitate the shortest possible
zero-dispersion wavelength while the fibre is still single-mode
for longer wavelengths, i.e. $\lambda_c\leq \lambda_0$.

\begin{figure}[b]
\begin{center}
\epsfig{file=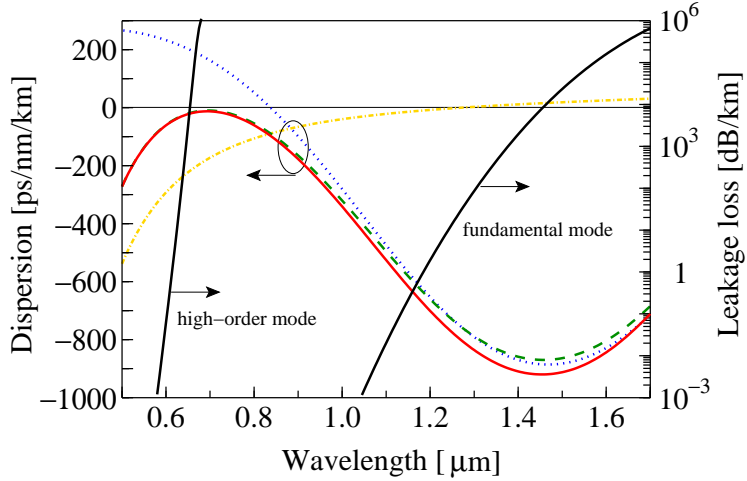, width=0.75\textwidth,clip}
\end{center}
\caption{Left axis: Dispersion parameter versus wavelength for a
photonic crystal fibre with $d/\Lambda=0.7$, $\Lambda=700\,{\rm
nm}$, and $N=10$. The solid curve shows the total dispersion,
Eq.~(\ref{eq:D}), obtained through a self-consistent numerical
solution of the wave equation, Eq.~(\ref{eq:waveequation}), while
the dashed line shows the approximate addition of waveguide and
material dispersion, Eq.~(\ref{eq:Dapprox}). The dotted curve
shows the pure waveguide contribution, Eq.~(\ref{eq:Dw}), while
the dot-dashed curve shows the silica material dispersion,
Eq.~(\ref{eq:Dm}). Right axis: Leakage loss versus wavelength for
the fundamental and the first high-order mode. The rapid increase
in the leakage loss of the high-order mode marks the cut-off
wavelength. } \label{fig2}
\end{figure}

\begin{figure}[t!]
\begin{center}
\epsfig{file=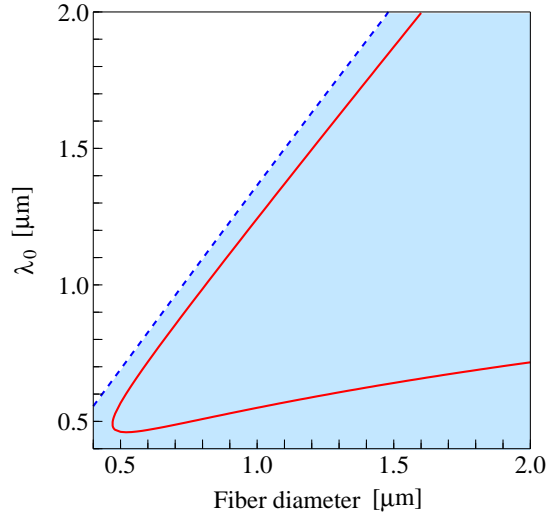, width=0.55\textwidth,clip}
\end{center}
\caption{Zero-dispersion wavelength versus fibre diameter for a
strand of silica in air. The dashed line indicates the cut-off
wavelength and the blue shading indicates the multi-mode phase. }
\label{fig3}
\end{figure}

\section{Competition between material and wave-guide dispersion}

The linear dynamics and response of a waveguide is typically
studied within the framework of temporal harmonic modes,
$\vec{E}(\vec{r},t)=\vec{E}(\vec{r}_\perp)e^{-i\omega t}=\vec{\cal
E }(\vec{r}_\perp)e^{i(\beta z -\omega t)}$, where
$\vec{E}(\vec{r})$ is a solution to the vectorial wave equation

\begin{equation}\label{eq:waveequation}
\nabla\times\nabla\times \vec{E}(\vec{r}) =
\varepsilon(\vec{r},\omega) \frac{\omega^2}{c^2}\vec{E}(\vec{r}).
\end{equation}
Here, $\varepsilon(\vec{r},\omega)=n^2(\vec{r},\omega)$ is the
spatially dependent dielectric function of the composite
air-silica dielectric medium, see Fig.~\ref{fig1}, and
$n(\vec{r},\omega)$ is the corresponding refractive index.

Solving the wave equation, Eq.~(\ref{eq:waveequation}), provides us
with the dispersion relation $\omega(\beta)$ which contains all
information on the spatial-temporal linear dynamics of wave
packets and it is furthermore important input for studies of
non-linear dynamics. However, quite often the dynamics and
evolution of pulses are quantified by the derived dispersion
parameters $\beta_n=\left(\partial^n
\omega/\partial^n\beta\right)^{-1}$ with $\beta_1=v_g^{-1}$ being
recognized as the inverse group velocity. The group-velocity
dispersion $\beta_2$ is in fiber optics commonly quantified by the
dispersion parameter $D$ (typically in units of ps/nm/km) which
can be written in a variety of ways including
\begin{equation}\label{eq:D}
D= \frac{\partial^2
\beta}{\partial\lambda\partial\omega}=\frac{\partial\beta_1}{\partial
\lambda} =-\frac{\omega^2}{2\pi c}\beta_2
\end{equation}
where $\lambda=2\pi c/\omega$ is the free-space wavelength. The
dispersion in the group velocity makes different components in a
pulse propagate at different speeds and thus pulses will compress
or broaden in time depending on the sign of $D$. The
zero-dispersion wavelength $\lambda_0$, defined by
$D(\lambda_0)=0$, is thus of particular relevance to pulse
dynamics in general and non-linear super-continuum generation in
particular. By pumping the fiber with high-intensive ultra-short
nano and femto-second pulses near $\lambda_0$ the pump pulses will
loosely speaking maintain their high intensity for a longer time
(or propagation distance) thus allowing for pronounced non-linear
interactions.

The finite group-velocity dispersion is a consequence of both the
dispersive properties of the host material itself as well as the
strong transverse localization of the light caused by the
pronounced spatial variations in the dielectric function. Since
the variation with frequency of the refractive index of silica is
modest (at least in the transparent part of the spectrum) the
dielectric function satisfies

\begin{eqnarray}
\varepsilon(\vec{r}_\perp,\omega)&=\varepsilon_0(\vec{r}_\perp,\tilde\omega)+\delta\varepsilon(\vec{r}_\perp,\omega),\\
\delta\varepsilon(\vec{r}_\perp,\omega) &=
\varepsilon(\vec{r}_\perp,\omega) -
\varepsilon_0(\vec{r}_\perp,\tilde\omega)
\ll\varepsilon_0(\vec{r}_\perp,\omega)
\end{eqnarray}
where $\tilde\omega$ is some arbitrary, but fixed frequency where
consequently $\delta\varepsilon=0$. The high attention to the
telecommunication band has often made $\tilde\lambda=1550\,{\rm
nm}$ a typical choice, though this is by no means a unique choice.
From $\varepsilon_0$ one may define a pure waveguide contribution
$D_w$ to the dispersion parameter by the definition

\begin{equation}\label{eq:Dw}
D_w= \frac{\partial^2 \beta_0}{\partial\lambda\partial\omega}
\end{equation}
where $\beta_0(\omega)$ is the solution to the wave equation with
$\varepsilon(\vec{r}_\perp,\omega)=\varepsilon_0(\vec{r}_\perp,\tilde\omega)$,
i.e. the frequency dependence of the dielectric function is
ignored. Furthermore, this has the consequence that the wave
equation, Eq.~(\ref{eq:waveequation}), becomes scale invariant
which is very convenient from a numerical point of view since
results for one characteristic length scale can easily be scaled
to a different length scale.

For the dielectric material itself one likewise defines a material
dispersion $D^{{}}_m$ by solving the wave equation with
$\varepsilon(\vec{r}_\perp,\omega)=\varepsilon_m(\omega)=n_m^2(\omega)$.
From the simple homogeneous-space dispersion relation it readily
follows that

\begin{equation}\label{eq:Dm}
D^{{}}_m= -\frac{\lambda}{c}\frac{\partial^2
n^{{}}_m}{\partial\lambda^2}.
\end{equation}
Intuitively, one might speculate that the two kinds of sources of
dispersion simply add up and actually the approximation
\begin{equation}\label{eq:Dapprox}
D\approx D_w+D_m
\end{equation}
is used widely in the literature. While the approximation is
useful in qualitatively understanding the zero-dispersion
properties it is however also clear (see e.g. the work of Ferrando
{\it et al.}~\cite{Ferrando:2000}) that quantitative correct
results requires either a self-consistent solution of the wave
equation or some accurate perturbative
method~\cite{Laegsgaard:2003a,Laegsgaard:2003}.

In this paper we use a fully self-consistent solution of the wave
equation, Eq.~(\ref{eq:waveequation}). For the dielectric function
we use the frequency-independent value $\varepsilon=1$ in the
air-hole regions while we for silica employ the usual three-term
Sellmeier polynomial description,
\begin{equation}
\varepsilon_m(\lambda)=n_m^2(\lambda)=1+ \sum_{j=1}^3 \frac{a_j
\lambda^2}{\lambda^2-\lambda_j^2}
\end{equation}
where the absorption lines $\lambda_j$ and the corresponding
strengths $a_j$ are given by
\begin{eqnarray}
\lambda_1&=0.0684043\,{\rm \mu m},\hspace{1cm} &a_1=0.6961663,\\
\lambda_2&=0.1162414\,{\rm \mu m}, &a_2=0.4079426,\\
\lambda_3&=9.896161\,{\rm \mu m}, &a_3=0.8974794.
\end{eqnarray}

Figure~\ref{fig2} illustrates the typical dispersion properties of
a photonic crystal fibre. The strongly negative material
dispersion $D_m$ of silica below $\lambda\sim 1.3\,{\rm \mu m}$
tend to make the total dispersion $D$ of standard fibres negative
for $\lambda \lesssim 1.3\,{\rm \mu m}$, simply because of the
very weak waveguide contribution $D_w$. However, photonic crystal
fibres are contrary to this since the composite air-silica
cladding is seen to provide the guided mode with a strongly
positive waveguide dispersion $D_w$ which tends to shift the
zero-dispersion wavelength $\lambda_0$ far below $1.3\,{\rm \mu
m}$ towards the visible. While the material dispersion is fixed
the waveguide dispersion varies strongly in the phase-space
spanned by $d$ and $\Lambda$ and in this way the competition
between waveguide and material dispersion becomes a powerful
mechanism in engineering the zero-dispersion wavelength.

\section{A strand of silica in air --- the ultimate limit?}

As mentioned in the introduction the zero-dispersion wavelength
may be pushed to lower values by simply increasing the air hole
diameter and decreasing the pitch. For $d/\Lambda\rightarrow 1$
this will to some extend effectively leave us with a single strand
of silica surrounded by air. This limiting case has been
emphasized previously in the literature and in Fig.~\ref{fig3} we
reproduce the zero-dispersion wavelength results reported by
Knight {\it et al.}~\cite{Knight:2000}. Generally, the strand of
silica will have either two dispersion zeros or none, with the
exception of the special case where it only supports a single
dispersion zero. PCFs turn out to follow the same overall pattern
and the existence of two dispersion zeros turns out to have
interesting applications in super-continuum
generation~\cite{Andersen:2004}.

The results in Fig.~\ref{fig3} leave promises for a
zero-dispersion wavelength down to below 500~nm which will
eventually also be the ultimate limit for silica based PCFs.
However, as also indicated by the dashed line the large index
contrast between air and silica in general prevents single-mode
operation at the zero-dispersion wavelength. In the following we
will study to which degree the photonic crystal cladding concept
of PCFs can be used to circumvent this problem.

\section{Photonic crystal cladding as a modal sieve}

As demonstrated already by Birks {\it et al.}~\cite{Birks:1997}
the photonic crystal cladding of a PCF acts as modal sieve which
may prevent localization of high-order modes to the core region.
This so-called endlessly single-mode property has later been
studied in great detail
\cite{Mortensen:2002,Kuhlmey:2002,Mortensen:2003,Folkenberg:2003}
and it was recently argued that the endlessly single-mode
phenomena is a pure geometrical effect and that the PCF is
endlessly single mode for $d/\Lambda\lesssim 0.42$ irrespectively
of the fibre material refractive index~\cite{Mortensen:2005}. The
photonic crystal cladding thus serves to limit the number of
guided modes and at the same time the guided modes will to some
extend inherit the chromatic dispersion properties observed for
the strand of silica in air~\cite{Knight:2000}.

\begin{figure}[b!]
\begin{center}
\epsfig{file=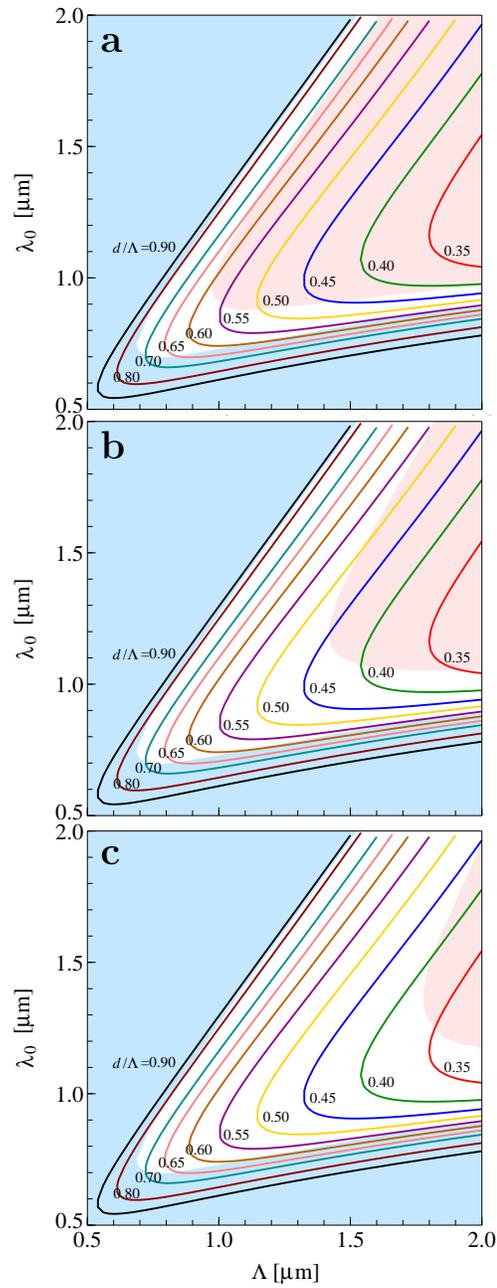,width=0.50\textwidth,clip}
\end{center}
\caption{Zero-dispersion wavelength $\lambda_0$ versus pitch
$\Lambda$ for different values of the normalized air-hole diameter
$d/\Lambda$. Panels a), b), and c) are for $N=6, 8$, and 10 rings
of air holes, respectively. Regions with a leakage loss larger
than 0.1 dB/km are indicated by red shading. Similarly, the
multi-mode regime is indicated by blue shading.} \label{fig4}
\end{figure}

The problem of zero-dispersion wavelength versus pitch has
previously been studied for PCFs with an infinite photonic crystal
cladding~\cite{Laegsgaard:2003} demonstrating curves qualitatively
resembling the curve in Fig.~\ref{fig3}. Here, we extend that work
to PCFs with a photonic crystal cladding of finite spatial extent.
In particular, we study the effect of a varying number $N$ of
rings of air holes surrounding the core region. We also study the
cut-off and leakage properties to explore the possibility for a
single-mode PCF with a zero-dispersion wavelength in the visible.

Our numerical solutions of the wave equation,
Eq.~(\ref{eq:waveequation}), are based on a finite-element
approach which is described in detail in Ref.~\cite{Koshiba:2002}.
For the calculation of the cut-off wavelength and the leakage loss
we refer to Refs.~\cite{Saitoh:2005,Koshiba:2005} and references
therein.

The results of extensive numerical simulations are summarized in
Fig.~\ref{fig4}. First of all we notice that the zero-dispersion
wavelength versus pitch has a curve-shape qualitatively resembling
the result in Fig.~\ref{fig3} for a strand of silica in air.
Furthermore, we see that the number $N$ of rings of air holes has
little influence on the zero-dispersion wavelength. In particular,
the results for $N=10$ are in full quantitative agreement with
those reported in Ref.~\cite{Laegsgaard:2003}. On the other hand,
the spatial extent $N\times\Lambda$ of the photonic crystal
cladding is as expected seen to have a huge impact on the leakage
loss~\cite{White:2001,Koshiba:2005} as seen from the red shading
indicating the region with a leakage loss exceeding 0.1~dB/km.
Furthermore, $N$ has as expected little effect on the cut-off
wavelength since the cut-off and the modal sieving is governed by
the width ($\Lambda-d$) of the silica regions between the air
holes~\cite{Mortensen:2005} rather than the spatial extent
$N\times\Lambda$ of the photonic crystal cladding.

Finally, we note that by choosing $d/\Lambda \sim 0.7$ we may
realize a PCF with a single zero-dispersion wavelength down to
$\sim 700$~nm with the fibre being single-mode for longer
wavelengths. We believe this to be the ultimate limit for
silica-based PCFs having a photonic crystal cladding comprising a
triangular arrangement of circular air holes. Such results have
been demonstrated experimentally by e.g. Knight {\it et
al.}~\cite{Knight:2000}. In practice, the limit might be pushed
slightly further toward the visible since real PCFs tend to have a
slightly shorter cut-off wavelength compared to the expectations
based on the ideal fibre structure~\cite{Folkenberg:2003}. Most
likely, this tendency originates in the presence of scattering
loss in real fibres which also acts in suppressing the high-order
modes even though they are weakly guided by the photonic crystal
cladding. In order to push the zero-dispersion wavelength further
into the visible one would have to tolerate guidance of high-order
modes or alternatively employ somewhat more complicated designs
involving a varying air-hole diameter throughout the
cladding~\cite{Jacobsen:2004}.

\section{Conclusion}

In conclusion we have studied the zero-dispersion wavelength
$\lambda_0$ in silica based photonic crystal fibres with special
emphasis on the interplay with the cut-off wavelength and leakage
loss. In the large parameter space spanned by the air-hole
diameter $d$ and the pitch $\Lambda$ we have identified the values
facilitating the shortest possible zero-dispersion wavelength
($\sim 700$~nm) while the fibre is still single-mode for longer
wavelengths.

We believe that our $\lambda_0$-maps are an important input for
the efforts in designing nonlinear photonic crystal fibres with
still shorter zero-dispersion wavelengths for super-continuum
generation in the visible.

\section{Acknowledgments}

N.~A~.~M. acknowledges discussions with J. L{\ae}gsgaard as well
as the collaboration on the zero-dispersion results in
Ref.~\cite{Laegsgaard:2003} which strongly stimulated the present
work.

\newpage


\end{document}